\documentclass{svproc}
\usepackage{graphicx}

\usepackage{url}

\begin{document}
\mainmatter              
\title{Equation of state of QCD matter within the Hagedorn bag-like model}
\titlerunning{QCD EoS within the Hagedorn bag-like model}  
%
\author{Volodymyr~Vovchenko\inst{1,2} \and Mark~I.~Gorenstein\inst{2,3} \and
Carsten~Greiner\inst{1} \and Horst~Stoecker\inst{1,2,4}}
\authorrunning{Volodymyr Vovchenko et al.} 
%
\tocauthor{Volodymyr Vovchenko, Mark I. Gorenstein, Carsten Greiner, and Horst Stoecker}
\institute{Institut f\"ur Theoretische Physik,
Goethe Universit\"at Frankfurt,\\
D-60438 Frankfurt am Main, Germany
\and
Frankfurt Institute for Advanced Studies, Giersch Science Center,\\
D-60438 Frankfurt am Main, Germany
\and
Bogolyubov Institute for Theoretical Physics, 03680 Kyiv, Ukraine
\and
GSI Helmholtzzentrum f\"ur Schwerionenforschung GmbH,\\ 
D-64291 Darmstadt, Germany}

\maketitle              

\begin{abstract}
The QCD equation of state at finite temperature and densities of conserved charges 
is considered in the
framework of a Hagedorn bag-like model, incorporating both the finite sizes of hadrons as well as their exponential mass spectrum. 
Augmented with non-zero masses of quarks and gluons in the bag spectrum, the model yields a fair quantitative description of lattice data on thermodynamic functions, the conserved charges susceptibilities, and Fourier coefficients of net-baryon density.
Both at zero and finite baryon densities a broad crossover transition between hadronic and quark-gluon matter is observed.
The model thus provides a thermodynamically consistent construction of a crossover equation of state for finite baryon number, electric charge and strangeness chemical potentials, which can be used in fluid dynamical simulations of heavy-ion collisions.


\end{abstract}

\section{Introduction}

The known hadron mass spectrum~\cite{Patrignani:2016xqp} is consistent with an exponential increase at large masses, as first suggested by Hagedorn long time ago~\cite{Hagedorn:1965st}. 
Other evidence for Hagedorn states comes from analyses of QCD thermodynamics~\cite{Majumder:2010ik,Lo:2015cca} and transport properties~\cite{NoronhaHostler:2008ju} in the vicinity of the chiral pseudocritical temperature $T_{pc}$, as well as from their role as a possible explanation of fast equilibration of hadrons in heavy-ion collisions~\cite{NoronhaHostler:2007jf,NoronhaHostler:2009cf}.
The Hagedorn states have recently been implemented into transport codes, as an alternative to the string mechanism of hadron production~\cite{Beitel:2016ghw,Gallmeister:2017ths}.

The class of Hagedorn bag-like models for the QCD equation of state~(EoS) has been constructed starting from the early 80s~\cite{Gorenstein:1981fa}, offering perhaps the first example of a framework which realizes a transition between hadronic matter and QGP.
One crucial ingredient there is an exponential spectrum of Hagedorn states, realized in the framework of the MIT bag model~\cite{Chodos:1974je,Kapusta:1981ay}.
Coupled with an introduction of finite eigenvolumes~\cite{Baacke:1976jv,Hagedorn:1980kb}, this allows to obtain a first-order, second-order, or a crossover transition in the gas of spatially extended quark-gluon bags, with thermodynamic properties at high temperatures being similar to the MIT bag model equation of state~\cite{Gorenstein:1998am,Gorenstein:2005rc,Zakout:2006zj,Zakout:2007nb}.
In contrast to many common constructions of the parton-hadron transition~(see e.g. \cite{Kolb:2000sd,1404.7540}), the transition in the Hagedorn model is described within a single partition function.

A comparison of the Hagedorn model crossover EoS with lattice QCD thermodynamics at $\mu_B = 0$ has first been considered in Ref.~\cite{Ferroni:2008ej}.
General qualitative features were found to be compatible with lattice QCD, although a quantitative description is lacking.
A more recent study~\cite{Vovchenko:2018eod} has achieved a semi-quantitative description of the lattice data 
by introducing quasiparticle-type quarks and gluons with non-zero masses into the bag spectrum.
This work reports on the recent advances in that direction, with a focus on finite chemical potentials of conserved charges.

\section{Model}

The Hagedorn bag-like model comprises of a statistical mechanics treatment of a multi-component system of non-overlapping particles which have finite sizes.
The particles carry integer values of the three QCD conserved charges: baryon number, electric charge, and strangeness.
The number of particle species with a particular mass $m$, eigenvolume $v$, and conserved charges $b$, $q$, and $s$ is characterized by a mass-volume density $\rho(m,v;b,q,s)$.
A generalized mass-volume density of states which depends on the chemical potentials is given by a Fourier transform of $\rho(m,v;b,q,s)$:
\begin{eqnarray}
\label{eq:rhofull}
\rho(m,v; \mu_B,\mu_Q,\mu_S) = \sum_{b,q,s = -\infty}^{\infty} 
e^\frac{b \mu_B}{T} \, e^\frac{q \mu_Q}{T} \, e^\frac{s \mu_S}{T}
\rho(m,v;b,q,s).
\end{eqnarray}

The density of states consists of two contributions, $\rho = \rho_H + \rho_Q$.
Here
\begin{eqnarray}
\label{eq:rhoH}
\rho_H(m,v;\mu_B,\mu_Q,\mu_S) = \sum_{i \in {\small \textrm{pdg}}} 
e^\frac{b_i \mu_B}{T} \, e^\frac{q_i \mu_Q}{T} \, e^\frac{s_i \mu_S}{T}
\,  d_i \, \delta(m - m_i) \, \delta(v - v_i). 
\end{eqnarray}
is the discrete part of the particle spectrum, consisting of the low-mass, established hadrons and resonances listed in Particle Data Tables~\cite{Patrignani:2016xqp}. $m_i$ and $d_i$ are the $i$th species' mass and degeneracy, respectively, and $v_i = m_i / (4B)$ is its bag-model motivated eigenvolume. 
$B$ is the bag constant, which will be specified later.

The continuum part of the particle spectrum is $\rho_Q$. It corresponds to the exponential spectrum of quark-gluon bags, evaluated assuming bags filled with non-interacting quarks and gluons with constant masses~\cite{Kapusta:1981ay,Gorenstein:1982ua,Vovchenko:2018eod}:
\begin{eqnarray}
\label{eq:rhoQ}
\rho_Q(m,v; \mu_B,\mu_Q,\mu_S) & = C \, v^{\gamma} \, (m - Bv)^{\delta} \, \exp \left\{ \frac{4}{3} [\sigma_{Q}(\mu_B,\mu_Q,\mu_S) \, v]^{1 \over 4} \, 
(m-Bv)^{3 \over 4} \right\} \nonumber \\ 
& \quad \times \theta(v-V_0) \, \theta(m - Bv - M_0).
\end{eqnarray}
Here $\sigma_{Q}(\mu_B,\mu_Q,\mu_S)$ is three times the pressure of an ideal gas of massive quarks and gluons.
$C$, $\gamma$, $\delta$, $V_0$ and $M_0$ are model parameters.
The values of the exponents $\gamma$ and $\delta$ determine the nature of the transition between hadrons and QGP.
The different possibilities were categorized in Ref.~\cite{Begun:2009an}.
In the following we only consider the crossover case.

To evaluate the system pressure one needs to incorporate the eigenvolume effects.
This is achieved through the isobaric ensemble~\cite{Gorenstein:1981fa}.
For the case of a crossover transition the resulting pressure is given by a transcendental equation~\cite{Vovchenko:2018eod}:
\begin{eqnarray}
\label{eq:pfull}
& & p(T,\mu_B,\mu_Q,\mu_S) =
T \sum_{i \in \rm pdg} d_i \, \phi(T,m) \, \exp\left( \frac{b_i \mu_B + q_i \mu_Q + s_i \mu_S}{T} \right) \, \exp\left(- \frac{m_i p}{4 B T} \right) \nonumber \\
& & + \frac{C}{\pi} \, T^{5+4\delta} \, [\sigma_Q]^{\delta+{1\over2}} \, [B+\sigma_Q T^4]^{3 \over 2} \,
\left(\frac{T}{p-p_B}\right)^{\gamma+\delta+3} \, \Gamma\left[\gamma+\delta+3, \frac{(p-p_B)V_0}{T} \right].
\end{eqnarray}
Here $\Gamma$ is the incomplete Gamma function.
Equation~(\ref{eq:pfull}) is solved numerically.

\section{Results}

For calculations we use the following parameter set:
\begin{eqnarray}
& & \gamma = 0, \quad  \delta = -2,  \quad B^{1/4} = 250~\textrm{MeV}, \quad C = 0.03~\textrm{GeV}^{4}, \quad V_0 = 4~\textrm{fm}^3,
\\
& & m_u = m_d = 300~\textrm{MeV}, \qquad m_s = 350~\textrm{MeV}, \qquad m_g = 800~\textrm{MeV}.
\end{eqnarray}
These parameter values were obtained in Ref.~\cite{Vovchenko:2018eod} by constraining the model to lattice QCD thermodynamics at $\mu_B = 0$, yielding a reasonable description of the lattice data on thermodynamic functions, conserved charges susceptibilities, and Fourier coefficients of net-baryon density~(see Ref.~\cite{Vovchenko:2018eod} for the agreement level). 

\begin{figure}
\centering
\includegraphics[width=.49\textwidth]{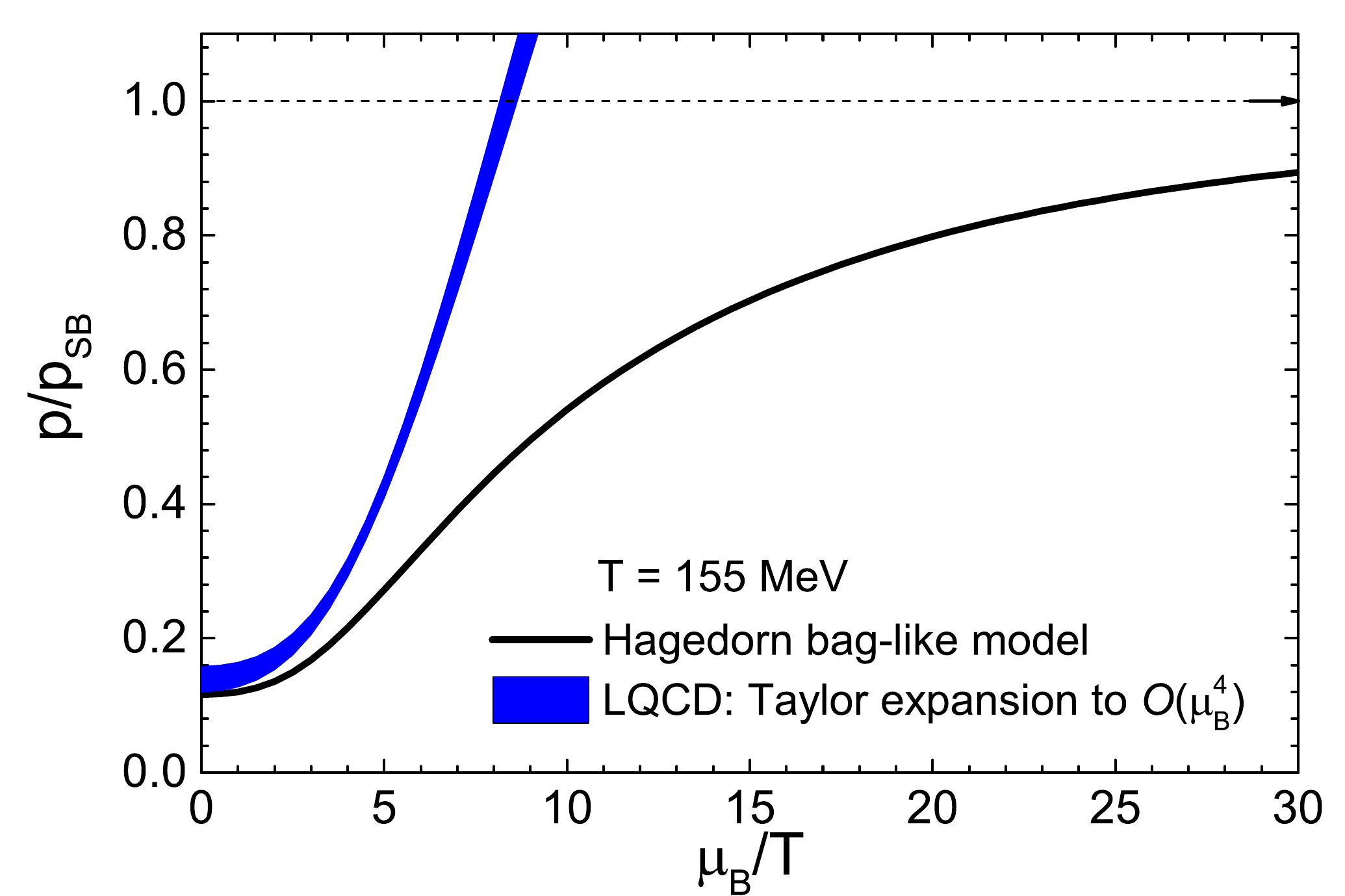}
\includegraphics[width=.49\textwidth]{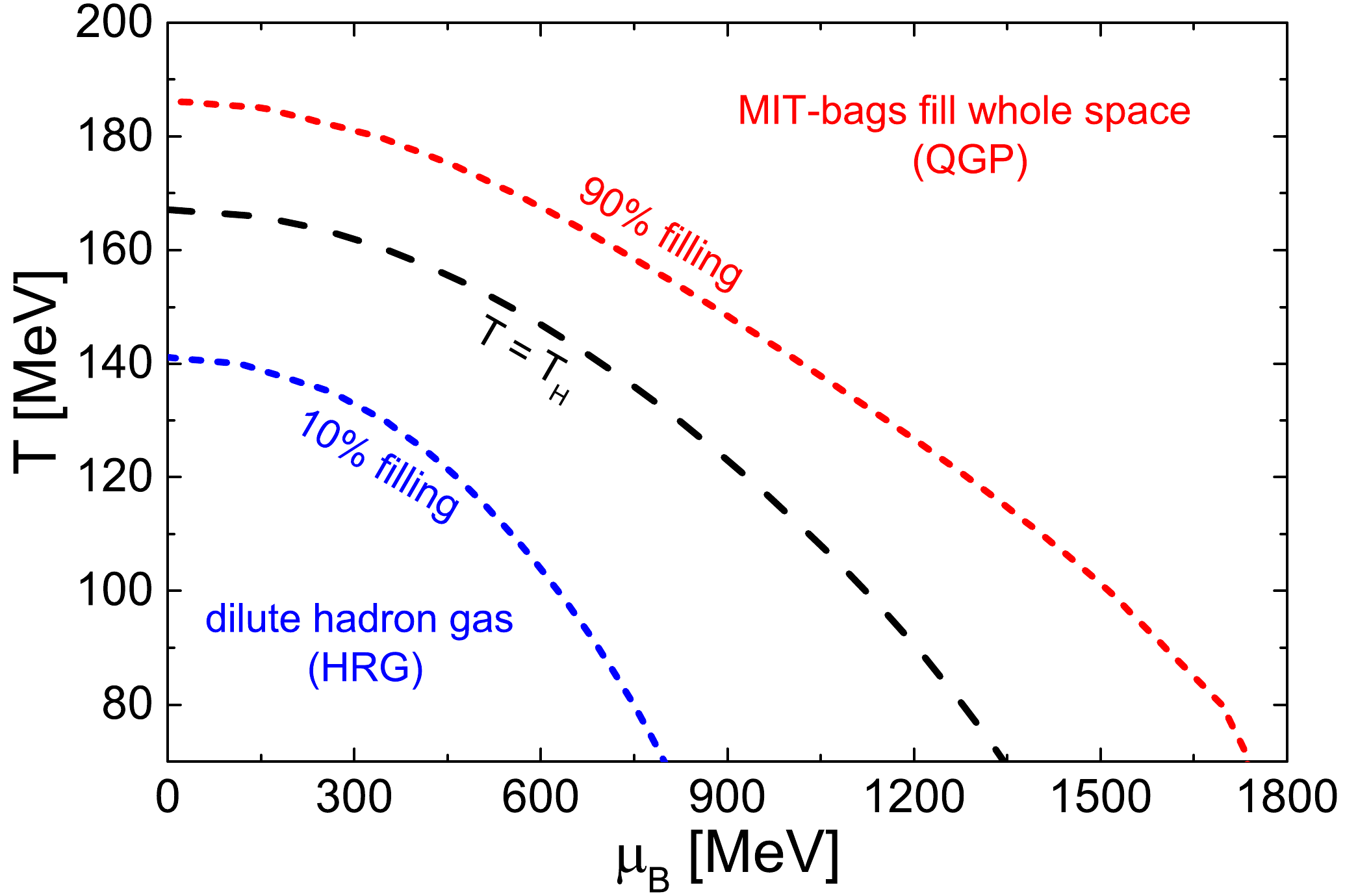}
\caption{\emph{Left panel:} Dependence of the pressure on the baryochemical potential along the $T = 155$~MeV isotherm.
The pressure is scaled by the corresponding value in the Stefan-Boltzmann limit of 3-flavor QCD. 
\emph{Right panel:} The phase structure of the Hagedorn bag-like model in the $\mu_B$-$T$ plane. The dashed black line corresponds to the $\mu_B$-dependent Hagedorn temperature 
whereas the blue and red dashed lines depict the contours corresponding to 10\% and 90\% eigenvolume filling fractions, respectively.}
\label{fig:finitemub}
\end{figure}

We emphasize now, that the present Hagedorn model is defined not only for zero chemical potentials, but also for finite values of $\mu_B$, $\mu_Q$, and $\mu_S$.
Here we explore the structure and behavior of the model at finite baryon density, something which was not done before in this framework quantitatively.
In the present work we let $\mu_Q = \mu_S = 0$ and perform calculations for finite $\mu_B$ values only.
The left panel of Fig.~\ref{fig:finitemub} depicts the $\mu_B/T$ dependence of the pressure along the $T = 155$~MeV isotherm. The pressure is scaled by the corresponding value in the Stefan-Boltzmann~(SB) limit of 3-flavor QCD -- the expected $\mu_B \to \infty$ limit.
The pressure of the Hagedorn model exhibits a consistent approach towards the SB limit at large $\mu_B/T$.
The behavior is smooth in the entire range of $\mu_B/T$ values considered.
For comparison we also depict the pressure by using a Taylor expansion truncated at $\mathcal{O}(\mu_B^4)$, with expansion coefficients taken from the lattice data~\cite{Bazavov:2017dus}.
The truncated Taylor expansion behaves similarly to the Hagedorn model at moderate $\mu_B/T < 3$, but becomes inaccurate at larger chemical potentials, overshooting the SB limit at $\mu_B/T \approx 8$.

To further clarify the phase structure of the Hagedorn model we consider the $\mu_B-T$ plane~(right panel of Fig.~\ref{fig:finitemub}).
The dashed black line depicts the $\mu_B$-dependent of the Hagedorn temperature of the quark-gluon bag spectrum in Eq.~(\ref{eq:rhoQ}).
This line, defined through a transcendental equation $T_H = [3B / \sigma_Q(T_H)]^{1/4}$, corresponds to the region of the phase diagram where heavy Hagedorn states start significantly contributing to the thermodynamics. The line can roughly be viewed as a crossover line for the transition between normal hadronic matter and the QGP. Another interesting quantity is the eigenvolume filling fraction which corresponds to the fraction of the whole space occupied by finite-sized particles.
The blue and red dashed lines in Fig.~\ref{fig:finitemub} depict the contours corresponding to 10\% and 90\% eigenvolume filling fractions, respectively.
The matter with less than 10\% filling is dilute and very similar to a hadron resonance gas.
In the opposite side of the phase diagram one has large filling fractions, meaning that most of the space is occupied by bags filled with QGP.
The results shown in Fig.~\ref{fig:finitemub} suggest that the transition between these two regimes is rather broad, not only at $\mu_B = 0$, but also in the baryon-rich region of the phase diagram.

\section{Summary}

The Hagedorn bag-like model with quasiparticle-type parton masses
provides a reasonable description of parton-hadron crossover at zero and finite densities of conserved charges within a single partition function. 
It is thus suitable as an input for fluid dynamical simulations of heavy-ion collisions.
One possible future extension is incorporating a hypothetical first-order phase transition in the baryon-rich region, e.g. as outlined in Ref.~\cite{Gorenstein:2005rc}.
Such a generalization can then be used to predict testable signatures of a first-order transition at finite baryon densities.

%
%

\end{document}